\documentclass[doublecol]{epl2} 
\usepackage{lineno,hyperref}
\usepackage{amsmath}

\title{Thermopower generation and thermoelectric cooling in a Kane-Mele normal-insulator-superconductor nano-junction}
\shorttitle{Title} %Insert here a short version of the title if it exceeds 70 characters

\author{Priyadarshini Kapri\inst{1} \and Saurabh Basu\inst{1} }
\institute{                    
  \inst{1} Indian Institute of Technology Guwahati, Assam-781039, India
}

\abstract{
We have studied thermoelectric effect of a Kane-Mele normal
- insulator - superconductor (KMNIS) junction at ultra-low temperatures 
using a modified version of the well-known Blonder-Tinkham-Klapwijk
(BTK) theory. Since both the (electronic) charge and thermal current due to the carriers are sensitive to the strengths of the spin-orbit coupling (SOC) present in the Kane-Mele model,
a tunability of this junction device with regard to its thermoelectric 
properties can be experimentally achieved by certain techniques that are used to
manipulate the values of the spin-orbit couplings. We have 
computed the Seebeck coefficient, the Figure of Merit, the thermoelectric cooling, the coefficient of performance of the KMNIS junction as a self-cooling device
 and investigated the role of the Rashba SOC (RSOC) and intrinsic SOC (ISOC) parameters therein.
Our results on the thermoelectric cooling indicate practical realizability and 
usefulness for efficient
cooling detectors, sensors, and quantum devices and hence could be crucial to
experimental success of the thermoelectric applications of such junction devices. Further we have briefly touched upon the condition that distinguishes
transmission through a topological insulator from an ordinary one.
}

\begin{document}

\maketitle

\section{Introduction}
One of the most attracted theoretical attention in the field of condensed matter physics since the 
several decades has been the study of the graphene \cite{Novoselov1,Martino}, a two dimensional single 
layer of hexagonal lattice of carbon atoms.
The conduction and the valance bands in graphene touch each other along the six Dirac points
 where the quasiparticle excitations show a linear Dirac-like energy dispersion. 
The unique geometrical structure of graphene has generated tremendous interest 
in different field of applications, such as, electronics  \cite{Novoselov1,Martino,Neto}, opto-electronics  \cite{Bonaccorso,Avouris}, 
and spintronics  \cite{Moghaddam,Pesin,Han,Ghosh}. Further, some other interesting phenomena such as, 
anomalous quantum Hall effect \cite{Zhang1,Novoselov3}, chiral tunneling  \cite{Katsnelson,Beenakker}, Klein paradox \cite{Katsnelson,Zareyan} have been obtained 
in graphene.  Moreover, the superconducting features can be induced in graphene 
by possible intercalation with dopant molecules  \cite{Uchoa} or via proximity effects  \cite{Buzdin,Heersche}. 
Such prospects provide a newer scopes of fabricating devices based 
on hybrid structures of graphene based superconductors.  

Recently, the thermal and thermoelectric properties of
 graphene structure have gained much attention because of the 
large Seebeck coefficient and high thermal conductivity obtained in graphene sheets \cite{Balandin,Zuev,Dragoman}. 
Previously, due to the experimental limitations in accessing nano-scale devices, 
the charge and spin dependent thermoelectric properties were often ignored \cite{Uchida,Jaworski}.
But recently improved techniques in low temperature measurements 
devices provides an opportunity of experimental 
observation of thermoelectric physics. Very recently Zuev et al \cite{Zuev} and Wei et al. \cite{Wei} have performed 
theoretical and experimental investigation of the thermoelectric effects of graphene sheets.

Two types of spin-orbit couplings (SOC) are proposed in graphene, 
the Rashba spin SOC and the intrinsic SOC where it has been predicted by Kane and Mele 
that the presence of both the spin-orbit term may be the reason behind realizing a new type of topological 
state which is known as the quantum spin Hall state (QSH)\cite{KM}. However owing to the very small strength 
of the spin gap  (typically $\sim 0.01-0.05 meV$) \cite{Min,Yao}  in graphene, the QSH state is not achieved in experiments.  
But there are possible methods available to induce enhanced SOC strengths in graphene, 
such as via adatoms \cite{Weeks}, using proximity effect of a three dimensional topological insulators  \cite{Kou,Zhang2},
 by functionalization with methyl groups \cite{Zollner} etc. It is experimentally obtained that in a graphene nano-sheet 
the RSOC strength can be enhanced up to $17 meV$ by proximity effects. 
Further, it is observed that from gold (Au) intercalation at the graphene-Ni interface  \cite{Marchenko}, the RSOC is enhanced up to approximately
$100 meV$.  A large
Rashba splitting about $225$ meV in epitaxial graphene layers on the surface of Ni  \cite{Dedkov} 
and a giant RSOC at the graphene-Ir surface from Pb intercalation have been observed \cite{Calleja}. 
Moreover the tunability
of RSOC strength via an external gate voltage provides an additional impetus
in the field of spintronics. It is worth to mention that SOCs are very significant
and hence cannot be skipped in the context of charge transport. 

In a parallel front, the quantum transport through the junction devices are gaining increased attention 
in the field of modern research for developing the nano-devices at atomic/molecular level. 
The junction devices have interesting applications in the fields of thermoelectric, thermometric, solid state cooling etc. 
In the past a good number of studies on junction has been performed in these fields \cite{Mastomaki,Liu,Nahum,Bardas,Wy1,Wy2}
 and the junction devices are very useful in wide range of experiments and applications \cite{Feshchenko,Clark,Miller}.
The recent development in the field of thermoelectric physics in small scale junction 
devices provides a new direction for fabricating self-cooling devices, thermopower devices etc.

Motivated by the above, we have performed an extensive study of
 the thermoelectric effect of a Kane-Mele normal-insulator-superconductor (KMNIS)
 nano-junction by employing the modified Blonder-Tinkham-Klapwijk (BTK) theory
which describes the low energy transmission characteristics of nano/mesoscopic junctions.
Physically the scenario corresponds to adatom decorated graphene NIS junction to account for finite strengths of
the spin-orbit couplings. Thus a KMNIS is used interchangeably with adatom decorated graphene NIS junctions.
 We have computed the spin resolved thermopower, Figure of Merit, thermal current, coefficient of performance and 
explored how the spin-orbit couplings (induced by adatoms or otherwise) assume roles in shaping up 
the thermoelectric properties of such a junction.
%For better readability, let us comment on the organization of our work. To make notations clear, we briefly describe the system of our study and 
%then outline of the BTK theory to obtain the electrical charge current through our KMNIS in section(\ref{sec:2}). 
%Calculation of Seebeck coefficient are presented in subsection(\ref{subsec:2.1}) and  
%in subsection(\ref{subsec:2.2}) we depict the theory of thermoelectric cooling.
%The results on the thermopower and thermoelectric cooling of this junction device are discussed in section(\ref{sec:3}).
%In section(\ref{sec:4}) we summarize our results and highlight the tunability of the junction device with regard to its thermoelectric properties.
%========================================================================================================================
\section{Theoretical model: BTK approach}
\label{sec:2}
\begin{figure}[!h]
\centering
\includegraphics[scale=0.4]{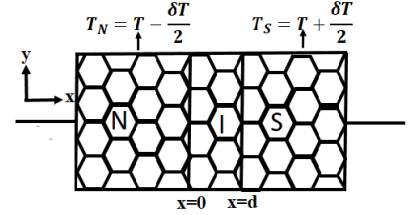}
\caption{The KMNIS junction setup.}
  \label{fig1}
\end{figure}

In Fig.(\ref{fig1}) the schematic diagram of the junction setup has been shown  where 
the left electrode is a normal ($x\le0$) and the right electrode ($x\ge d$) is 
a superconducting material with the insulating layer is extending from  $x=0$ to $x=d$.
$V_0$ is the potential across the insulating barrier region which can be tuned by an external gate voltage. 
It is considered that the $x\ge d$ region of the junction system has been produced by proximity effect by an external superconductor.

The effective Hamiltonian for graphene with both the spin-orbit couplings is given by,
\begin{eqnarray}
\label{eq1}
H&=&-t_{1}\sum_{<ij>}a^{\dag}_{i}b_{j}+i\lambda_{I}\sum_{\langle\langle ij\rangle\rangle}V_{ij}(a^{\dag}_{i}\sigma_{z}a_{j}\\\nonumber &+&b^{\dag}_{i}\sigma_{z}b_{j})
+i\lambda_{R}\sum_{<ij>}a^{\dag}_{i}(\hat{\sigma}\times \hat{d_{ij}})\cdot\hat{n}b_{j}+\lambda_{\nu}\sum_{i}a^{\dag}_{i}a_{i}\\\nonumber&-&\lambda_{\nu}\sum_{i}b^{\dag}_{i}b_{i}+h.c.
\end{eqnarray}
The first term denotes the nearest neighbour (NN) hopping with a hopping strength $t_{1}$, the second term denotes the intrinsic spin-orbit coupling given 
by the next-nearest neighbours (NNN) hopping with the intrinsic spin-orbit coupling (ISOC) strength $\lambda_{I}$. $\langle\langle ij \rangle\rangle$ represents the summation over NNN,
$V_{ij}=+(-)$ if the hopping is clock-(anti clock) wise. The third term indicates Rashba term given by the nearest neighbour (NN) hopping with the Rashba spin-orbit
coupling (RSOC) strength $\lambda_{R}$, where $\hat{d_{ij}}$ signifies the 
unit vector from site $i$ to $j$, $\sigma$ denotes Pauli matrices and  $\hat{n}=\hat{x}$ is unit 
vector along the interface normal. 
An on-site energy term is given by, $\lambda_{\nu}$ which is different for the two sites within the cell.
In particular  we have considered $"+"$ energy  for A site and  $"-"$ energy for B site. Operators $a^{\dag}_{i}$ ($b^{\dag}_{i}$)
denotes the creation and  $a_{i}$ ($b_{i}$) denotes the annihilation of  an electron at $R_{i}$ site of the A(B) sublattices. 
Various details related for the simplification of the Hamiltonian and the BTK formalism applied to the Kane Mele normal-insulator-superconductor nano-junctions appear elsewhere \cite{PKapri}.

It is important to note that though in the presence of the RSOC and ISOC,  
there are opening of gaps in the electronic dispersion (insulating regime) at the valley points ($K$ and $K'$ ), the incident particles will still be able 
to pass through the insulating gap under certain condition. This can be stated as the following. The values of the intrinsic and the Rashba coupling terms should
be such that, $\lambda_{I\nu}^{'2}+\lambda^{'2}_{R})<(E_{N}^{F}\pm E)^2$ where 
$\lambda^{'}_{I\nu}=\lambda_{\nu}-\sigma3\sqrt3\lambda_I$, $\lambda'_R=3\lambda_R/2$, 
$E_F^N$ is the Fermi energy and $E$ is the energy of the carriers. If this condition is violated then there will 
be no transmission.
In a sense this condition may be contemplated to distinguish the topological insulators from that of the ordinary insulators.
In this regard, the inset of Fig.(1) of Ref.(\cite{Kane2}) may be seen where a diamond shaped region in the $\lambda_R$ (in units of $\lambda_I$)
{\it vs}  $\lambda_{\nu}$ (scaled by $\lambda_{I}$) is identified as the topological phase
where the transport is possible only via the edge states, while the bulk remains insulating.
To compute transport properties using  BTK formalism, the above inequality yields a condition for transmission, which, when violated
ensures no conductance.

The expression for the charge current through the KMNIS junction using BTK theory can be found to have the following form,
\begin{eqnarray}
I_{NS_{\sigma}}(E_{F}^{N},T_{N},E_{F}^{S},T_{S})=eA_{r}v_{F}^{N}\int\int\tau_{\sigma}(E,\theta_{N1})\\\nonumber
[f^{N}(E_{f}^{N},T^{N})-f^{S}(E^{S}_{F},T_{S})]N(E)dE\cos_{\theta_{N1}}d\theta_{N1}
\label{eqch52}
\end{eqnarray}
where $N(E)$ denotes the density of states, $v_{F}^{N}$ is the Fermi velocity, 
$A_{r}$ is the area of contact, and $f^{N}$, $f^{S}$ are the Fermi
distribution functions for the normal and the superconducting leads respectively. 
$\tau_{\sigma}(E,\theta_{N1})$ is the transfer probability
where $\tau_{\sigma}(E,\theta_{N1})=1-|b_{\sigma}(E,\theta_{N1})|^2+|a_{\sigma}(E,\theta_{N1})|^2$, $a_{\sigma}$ is amplitude of Andreev reflection,
$b_{\sigma}$ is amplitude of normal reflection, $\theta_{N1}$ is the angle of incidence for electrons. 
The Fermi energy variation across the junction system is assumed as,
$E_{F}(x)=E_{F}^{N}\Theta(-x)+E_{F}^{I}\Theta(d-x)+E_{F}^{S}\Theta(x-d)$,
where $E^N_F$ and $E^S_F$ are the Fermi energies of the normal the superconducting leads.
$E^I_F$ is the Fermi energy of the insulating barrier which is defined by, $E^I_F=E^N_F+V_0$, where $V_0$
is an external gate voltage.
As the normal state resistance, $R_N$ is given by, $R_N =\frac{1}{2e^{2}N_{0}v_{F}^{N}A_{r}}$ ($2$ comes due to 
the spin degeneracy, $N_{0}$ denotes the density of states at Fermi level), the electrical charge current takes the following form,
\begin{eqnarray}
\label{eq3}
I_{NS_{\sigma}}(E_{F}^{N},T_{N},E_{F}^{S},T_{S})=\frac{1}{2eR_{N}N_{0}}\int\int\tau_{\sigma}(E,\theta_{N1})\\\nonumber
[f^{N}(E_{f}^{N},T^{N})-f^{S}(E^{S}_{F},T_{S})]N(E)dE\cos{\theta_{N1} d\theta_{N1}}
\end{eqnarray}
where the energy dependent quantity,
$N(E)=\frac{|E_F^N+E|W}{\pi\hbar v_F^N}$
is the number of transverse modes in a graphene sheet of width $W$ \cite{Beenakker2}.
%============================================================================================================================================
\subsection{Seebeck coefficient}
Here we present the theory to calculate the Seebeck coefficient.
A temperature difference between two dissimilar materials produces a voltage difference between the two substances and this phenomenon is known as Seebeck effect.
Seebeck coefficient is a measurement of the amount of potential induced in the device for unit temperature difference and defined by, $S=\delta V/\delta T$.

We consider the left and right electrodes serve as independent temperature reservoirs where 
the left electrode is fixed at temperature, $T^{N}=T-\delta T/2$ and the right electrode is fixed at 
temperature $T^{S}=T+\delta T/2$. The population of electrons in the left and the right lead is described by the 
Fermi-Dirac distribution function, $f^{N}$ and $f^{S}$ respectively where $E_{F}^{N}=E_{F}^{S}$ at zero external bias.

Let us now consider an extra infinitesimal current induced by an additional voltage, $\delta V$ and the temperature difference, $\delta T$ across
the junction in an open circuit. The current induced by $\delta T$ and $\delta V$ are given by,
$(dI)_{T}=I(E_{F}^{N},T^{N},E_{F}^{S}=E_F^{N},T^{S}=T^{N}+\delta T$) and $(dI)_{V}=I(E_{F}^{N},T^{N},E_{F}^{S}=E_{F}^{N}+e\delta V, T^{S}=T^{N})$.
Suppose that the current cannot flow in an open circuit, thus $(dI)_{T}$ counter balances $(dI)_{V}$. It allows us to write,
\begin{equation}
\label{eq4}
dI=(dI)_{T}+(dI)_{V}=0
\end{equation}
where the expression for the $(dI)_{T}$ and $(dI)_{V}$ can be obtained from Eqn.(\ref{eq3}).
Now first order expansion of Fermi-Dirac distribution function in $(dI)_T$ and $(dI)_{V}$ yields the expression for the spin dependent Seebeck coefficient,
\begin{equation}
\label{eq5}
S_{\sigma}=\frac{\delta V}{\delta T}=\frac{\int\int dE d\theta_{N1}\cos{\theta_{N1}} E(E_F^N+E)\tau_{\sigma}(E,\theta_{N1})\frac{\partial f}{\partial E}}
{e T \int\int dEd\theta_{N1}\cos{\theta_{N1}} (E_F^N+E)\tau_{\sigma}(E,\theta_{N1})\frac{\partial f}{\partial E}}
\end{equation}
while the energy is shifted by Fermi energy.
Now the charge and spin Seebeck coefficients
are usually defined by \cite{Swirkowicz}, 
\begin{eqnarray}
\label{eq6}
S_{ch}=\frac{1}{2}(S_{up}+S_{down})\hspace{0.05 in}; S_{sp}=\frac{1}{2}|S_{up}-S_{down}|
\end{eqnarray}
which can be computed from Eqn.(\ref{eq5}) for $\sigma=up/down$.

The efficiency of the device depends upon  a quantity called
as 'Figure of Merit' (FM).
To get a clear idea of the efficiency, one should compute the spin dependent
FM which is given by,
\begin{equation}
\label{eq7}
Z_{\sigma}T=\frac{S_{\sigma}^{2}G_{\sigma}}{K_{\sigma}}T
\end{equation}
where $S_{\sigma}$ is Seebeck coefficient, $G_{\sigma}$ is electrical conductance, $K_{\sigma}$ is 
thermal conductance, and $T$ is absolute temperature.
$G_{\sigma}$ can be calculated from the relation $G_{\sigma}=\frac{dI_{NS_{\sigma}}}{dV}$ and is given by the form,
\begin{equation} 
G_{\sigma}=\frac{1}{2eR_{N}E_{F}^{N}}\int \int\tau_{\sigma}(E,\theta_{N1})(-\frac{\partial f}{\partial E})](E+E_{F}^{N})dE \cos{\theta_{N1}}d\theta_{N1}
\end{equation}
The thermal conductance $K_{\sigma}$, can be calculated from the relationship $K_{\sigma}=\frac{dJ_{NS_{\sigma}}}{dT}$ where $J_{NS_{\sigma}}$ is the
thermal current flowing from the normal region to the superconducting region. 
In the next subsection we shall discuss how the thermal current and the thermal conductance can be calculated.  

In addition, the charge FM ($Z_{ch}T$) and spin  FM ($Z_{sp}T$) are defined as \cite{Swirkowicz,Hatami,Chen},
\begin{eqnarray}
\label{eq9}
Z_{ch}T=\frac{S_{ch}^{2}(G_{up}+G_{down})T}{K_{up}+K_{down}}\hspace{0.05 in}; Z_{sp}T=\frac{S_{ch}^{2}|G_{up}-G_{down}|T}{K_{up}+K_{down}}
\end{eqnarray}
%=========================================================================================================================
\subsection{Thermoelectric cooling}
The flow of electrons can also transport the thermal energy through the junction which is responsible for the thermal current.
The thermal current is defined as the rate at which the thermal energy flows from left lead to right lead.
 As said earlier,
the left electrode, that is the normal lead serves as the 
cold reservoir and the right one serves as hot reservoir. 
The junction device is connected to an external bias voltage, $V_B=(E^N_F-E^S_F)/e$,
which drives the electrons to flow from the normal lead  to the superconducting lead.
Thus the electron removes the heat energy from the normal lead and transfers it to the
superconducting lead which further makes the cold reservoir (normal) cool. 
The energy conservation allows us to write,
\begin{eqnarray}
J_{NS_{\sigma}}(E^N_F,T^N;E^S_F,T^S)+I_{NS_{\sigma}}(E^N_F,T^N;E^S_F,T^S)V_B\\\nonumber=J_{SN_{\sigma}}(E^N_F,T^N;E^S_F,T^S)
\label{eq12}
\end{eqnarray} 
The analogue between the electronic charge current and the electronic thermal current allows us to write 
the outbound energy flow rate from the normal lead to the superconducting lead as,
\begin{eqnarray}\label{eq10}
J_{NS_{\sigma}}=\frac{1}{2e^2 R_NE_F^N}\int \int(E-eV_B)(E+E_F^N)\tau'_{\sigma}(E,\theta_{N1})\\\nonumber
[f^{N}(E-eV_B,T^N)-f^S(E,T^S)]dE \cos{\theta_{N1}}d\theta_{N1}
\end{eqnarray}
Similarly, the reverse, that is the rate at which the superconducting lead receives the thermal energy is written as,
\begin{eqnarray}\label{eq11}
J_{SN_{\sigma}}=\frac{1}{2e^2 R_NE_F^N}\int \int E(E+E_F^N) \tau'_{\sigma}(E,\theta_{N1})\\\nonumber
[f^{N}(E-eV_B,T^N)-f^S(E,T^S)]dE \cos{\theta_{N1}}d\theta_{N1}
\end{eqnarray}
while energies are shifted by Fermi energy and $\tau'_{\sigma}$ is given by the form,
\begin{equation}
\tau'_{\sigma}(E,\theta_{N1})=1-|b_{\sigma}(E,\theta_{N1})|^2-|a_{\sigma}(E,\theta_{N1})|^2
\end{equation}
The thermal conductance, $K_{\sigma}$ can be calculated from $\frac{dJ_{NS_{\sigma}}}{dT}$.
This normal-insulator-superconductor (NIS) junction can be regarded as the electronic cooling device only when $J_{NS_{\sigma}}>0$, which 
implies that it is capable to remove the heat from the cold reservoir, thereby making it cooler.

The performance of this junction as a self-cooling device can be measured by the coefficient of the performance ($COP$) where $COP$ is defined as the ratio of the 
heat removed from the cold reservoir to the electrical power needed for driving the system.
The $COP$ for electronic thermal current, namely, $COP_{\sigma}$ is given by,
\begin{eqnarray}
\label{eq14}
COP_{\sigma}=\frac{J_{NS_{\sigma}}}{I_{NS_{\sigma}}V_{B}}=\frac{J_{NS_{\sigma}}}{J_{NS_{\sigma}}-J_{SN_{\sigma}}}
\end{eqnarray}

%%==========================================================================================================================================%%
%===============================================================================================

%===========================================================================================================================%
\section{ Results and Discussions}
\label{sec:3}

We have investigated the thermoelectric effect of a 
graphene based normal-insulator-superconductor (NIS) 
junction device in the presence of Rashba and intrinsic spin-orbit
couplings assumably induced by the transition metal adatoms, 
where the effects of the spin-orbit couplings are mimicked 
by the Kane-Mele model \cite{KM}. When adatoms are adsorbed 
by graphene, the electrons of the outer most 
shell of adatoms are distributed among the carbon 
atoms of graphene. This causes a net positive charge 
in the vicinity of the adatoms and to screen this 
charge, electrons starts to accumulate surrounding 
 the adatom. This electron cloud results in  an inhomogeneous 
electric field which cause enhanced spin-orbit couplings.

Here we include a note on the values of various parameters used 
for our numerical computation. To put things in perspective, we have considered some reasonable values of $\Delta_0$, 
for example, $\Delta_0\sim(10^{-3} - 10^{-2} )eV$. 
$E_{F}^{N}$ has been considered as $50\Delta_0$. 
The Fermi velocity and the hopping parameter $t_{1}$ can be 
calculated from the Fermi energy through the relation $E_{F}=\hbar v_F k_F$ with $v_F=\frac{3t_1a}{2\hbar}$. The strength of the 
NN hopping, $\lambda_{R}$  and the  NNN hopping, 
$\lambda_I$ are varied in the range [$0:0.17t_{1}$] and  [$0:0.02t_{1}$] respectively. 
 In our work, the strength 
of the $\lambda_{R}$ and $\lambda_{I}$ are taken as parameters and 
the thermoelectric effects are computed for various choices of these parameters.  
Physically, it implies decorating the graphene nanoribbon 
by adatoms which induce the SOC couplings. 
It is worth mentioning that, using different techniques,
such as using an external gate voltage or organic solvents etc.,
it is possible to enhance the SOC strengths up to a
couple of orders of magnitude.  Finally, the staggered 
term is taken as $\lambda_{\nu}=0.1t_{1}$.
The temperature difference, $\delta T$ that exists across 
the junction is taken as, $\delta T=0.02\Delta_{0}$, ensuring that the difference is indeed small.
%The value of $\Lambda$ for the phonon thermal current generally is of the order of
%$10^{−17} (W/K^4 )$ \cite{Liu}. But in our computation, we make it a dimensionless quantity,
%with $\hbar = 1$ and multiplying by the factor $\Delta_{0}^{2}$ which finally yields $\Lambda \sim .467$.

%==================================================================================================
\subsection{\label{subsec:3.1} Seebeck Coefficient and Figure of Merit}
\begin{figure}[!h]
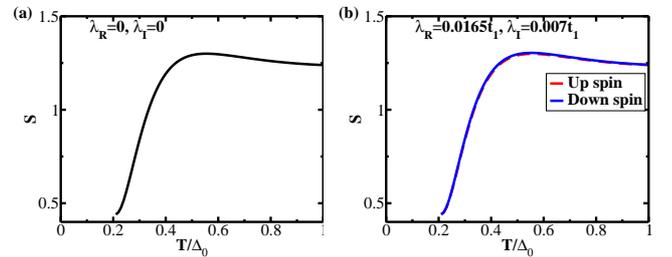

\centering
\includegraphics[scale=.16]{Fig2a.eps}\hspace{0.01in}
\includegraphics[scale=.16]{Fig2b.eps}
\caption{(Color online)
The variation of the Seebeck coefficient, $S$  as a function of temperature, $T$ (scaled by superconducting order parameter, $\Delta_0$) 
$(a)$ for pristine graphene,
$(b)$ for Au decorated graphene.}
\label{fig2}
\end{figure}

\begin{figure}[!h]
\centering
\includegraphics[scale=.17]{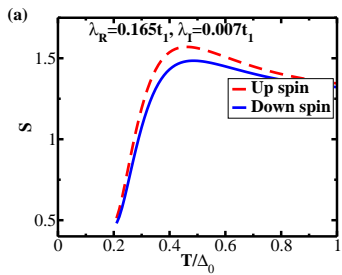}\hspace{0.05in}
\caption{(Color online)
The variation of the Seebeck coefficient, $S$  as a function of temperature, $T$ (scaled by superconducting order parameter, $\Delta_0$) 
for one order greater magnitude of RSOC parameter.}
%$$(b)$ for one order greater magnitude of ISOC parameter.}
\label{fig3}
\end{figure}

Initially we show the results for the Seebeck coefficient for a pristine graphene ($\lambda_R=\lambda_I=0$). 
The variation of  the Seebeck coefficient, $S$ as function of the temperature (in units of superconducting gap, $\Delta_{0}$) for a pristine graphene is shown in  Fig.(\ref{fig2}a).
The Seebeck coefficient, $S$ is dimensionless (since $e=1$ and $k_B=1$). 
To get an idea about the role of spin-orbit couplings on bahaviour of the thermopower, in Fig.(\ref{fig2}b) we present the variation of the 
spin resolved Seebeck coefficient, $S$ as function of  temperature for realistic values, that is, for Au decorated graphene which confirms from
first principal calculations, $\lambda_I=0.007t_1$ and
$\lambda_R=0.0165t_1$ \cite{Weeks}. A comparison between the two does not yield any significant change in the thermopower profile and neither one gets spin resolved signal.
 Thus, by some means, if we are able to enhance the SOCs by one
order of magnitude compared to the values available in the Au decorated graphene, there
could be noticeable effects of SOC. Thus in Fig.(\ref{fig3}) we have shown the
thermopower profile with RSOC strengths that are larger by one order of magnitude and indeed noticeable changes are obtained
with spin resolved contributions (up spin larger than that of the down spin).
Thus in the latter discussions we shall use these values of the SOCs strength.
In case of a normal junction (that is not a graphene based junction) in  presence of the Rashba term (the 
intrinsic term alone should not be responsible for a noticeable spin selection), there is no spin resolved thermopower.
It is clearly understood that with the inclusion of the SOC parameters, the Seebeck coefficient increases.

\begin{figure}[!h]
\centering
\includegraphics[scale=.17,{angle=270}]{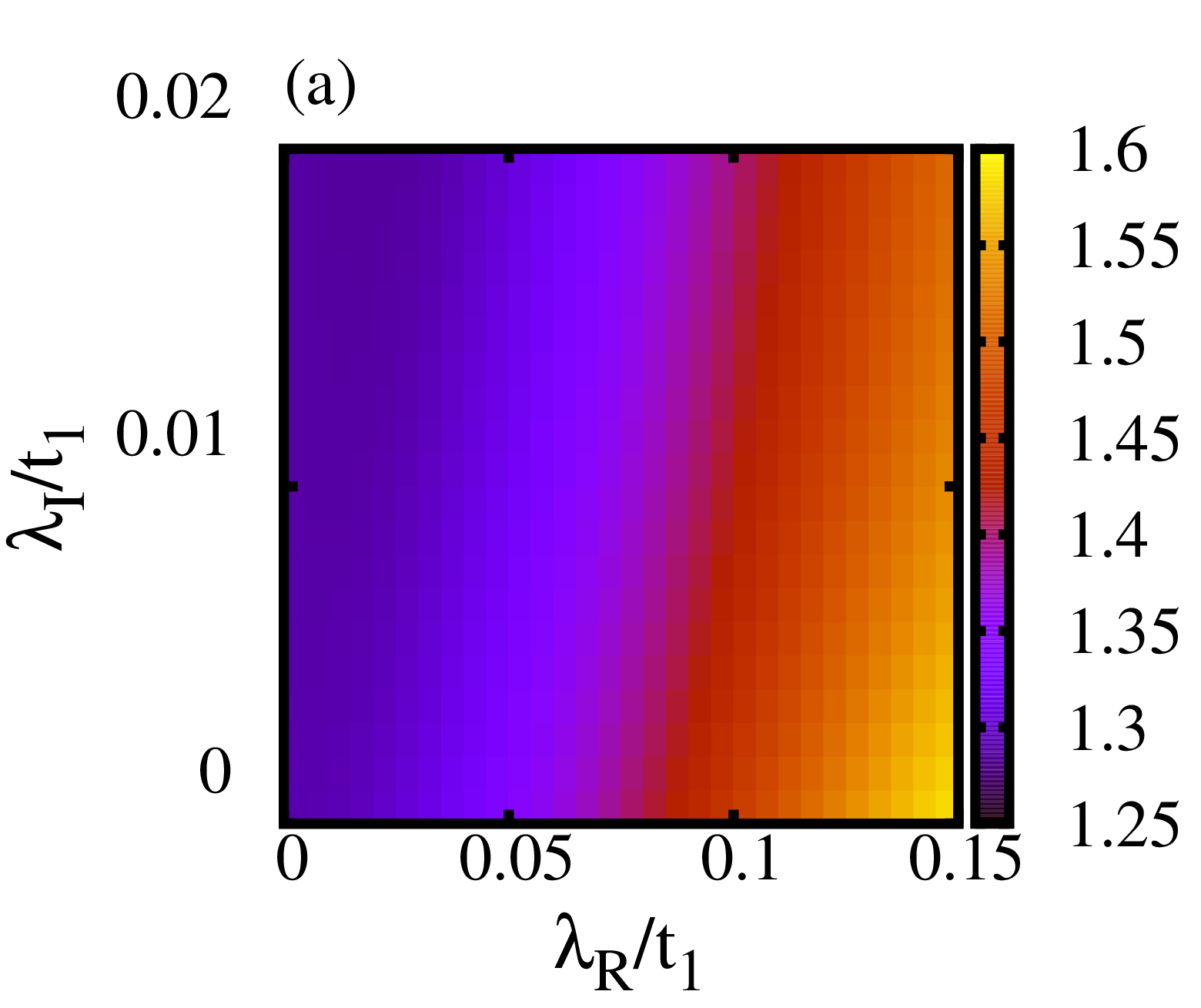}\hspace{0.01in}
\includegraphics[scale=.17,{angle=270}]{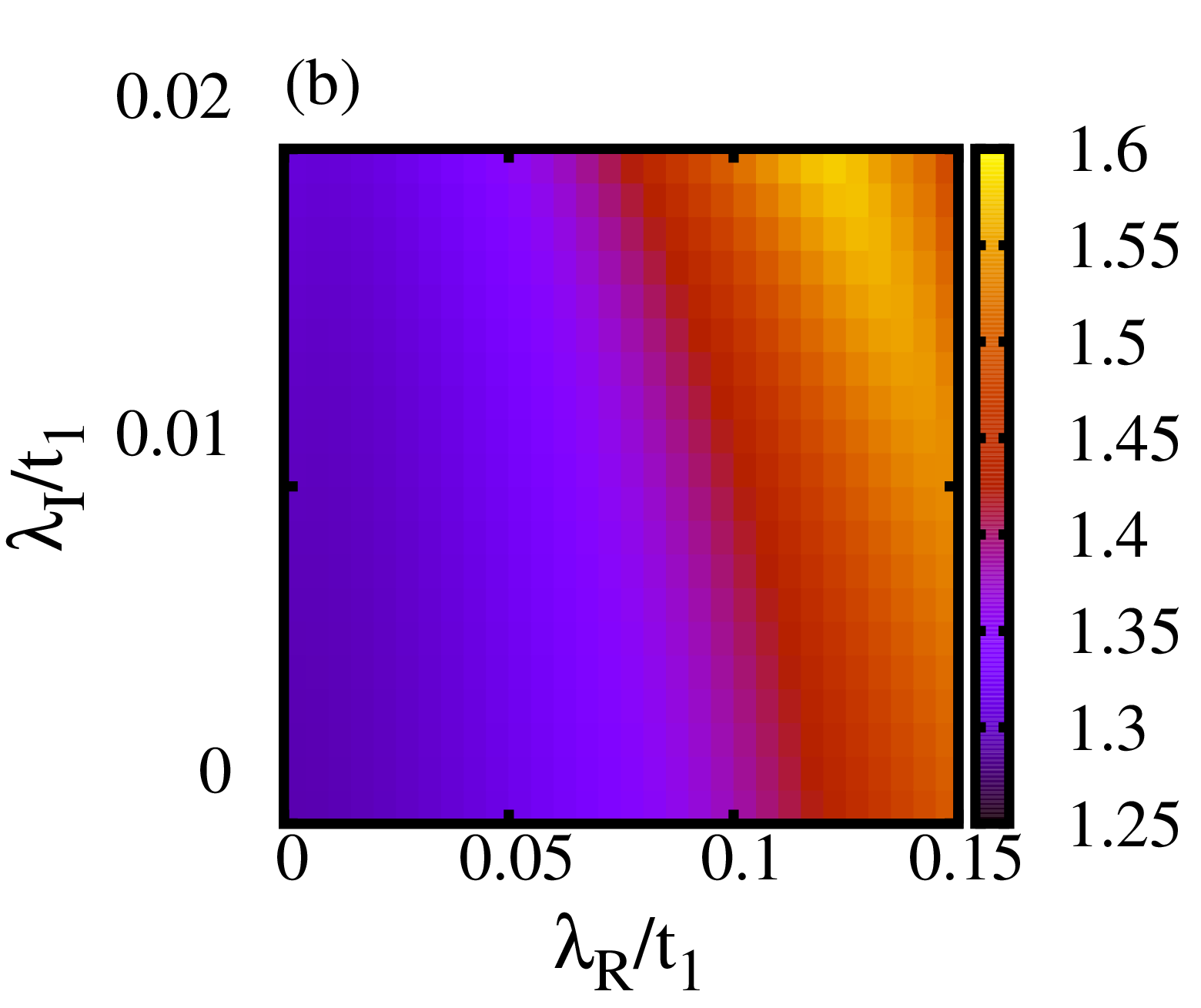}\\
\includegraphics[scale=.17,{angle=270}]{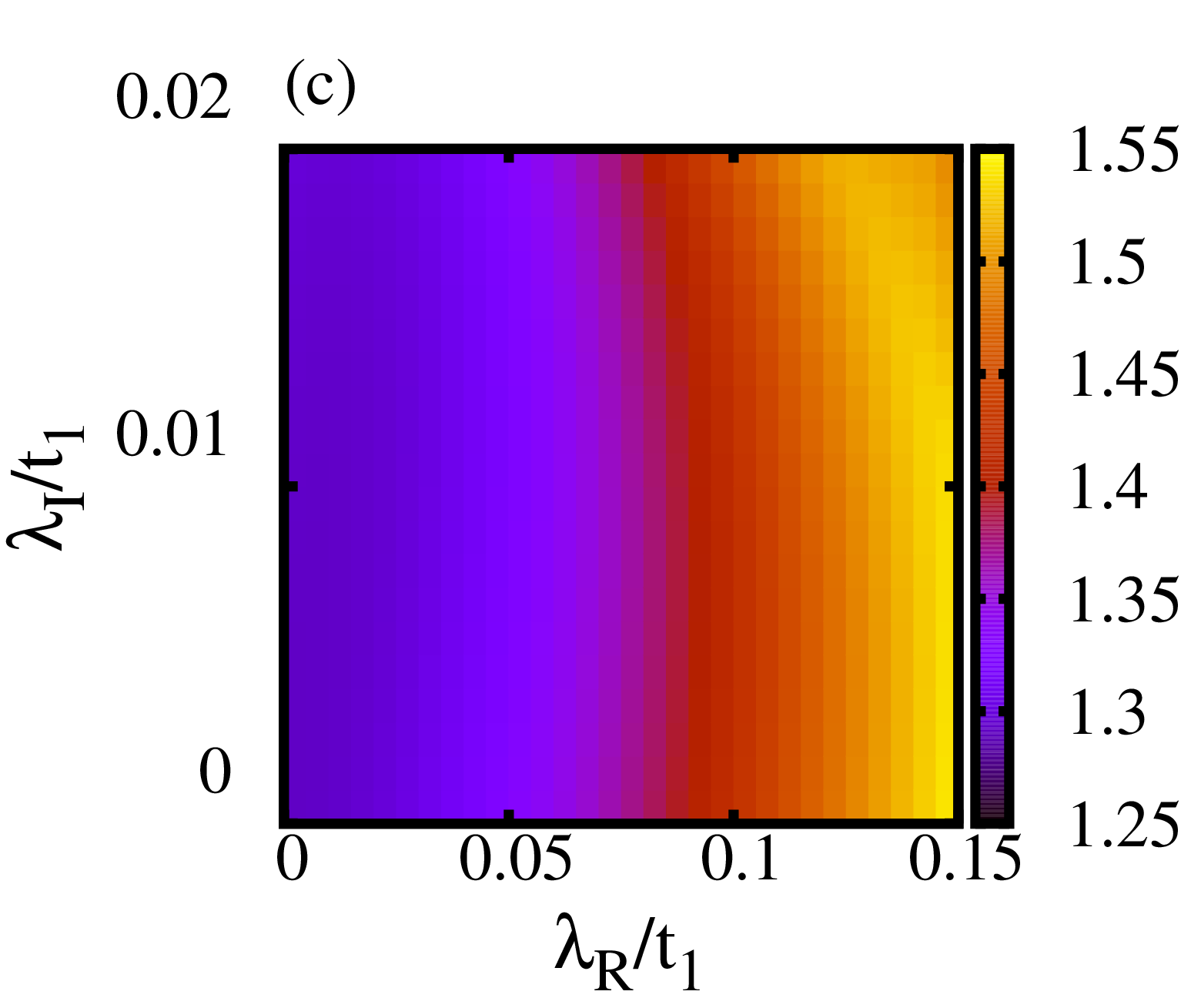}\hspace{0.01in}
\includegraphics[scale=.17,{angle=270}]{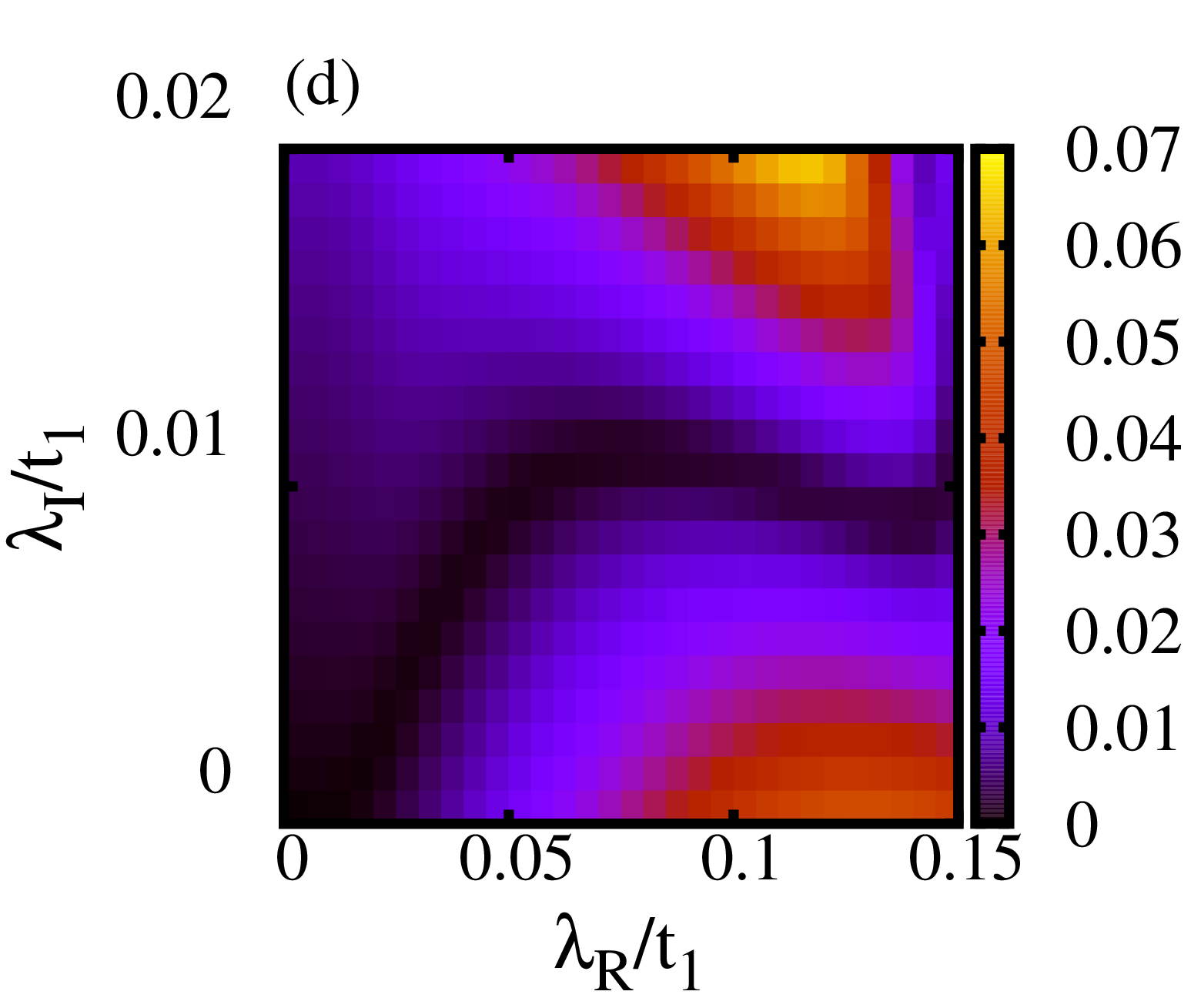}
\caption{(Color online)
The variation of the spin resolved Seebeck coefficient, $S$ as function of $\lambda_R$ and $\lambda_I$ (scaled by $t_1$) for
$(a)$ up spin,
$(b)$ down spin,
$(c)$ The variation of the charge Seebeck coefficient, $S_{ch}$ as function of $\lambda_R$ and $\lambda_I$ (scaled by $t_1$),
$(d)$ The variation of the spin Seebeck coefficient, $S_{sp}$ as function of $\lambda_R$ and $\lambda_I$ (scaled by $t_1$).}
\label{fig4}
\end{figure}

To get an idea how the spin resolved Seebeck coefficient vary with both of the spin-orbit 
couplings, and also to get an operating regime in the parameter space,
we have shown the spin resolved Seebeck coefficient as a function of $\lambda_R$ and 
$\lambda_I$ in Fig.(\ref{fig4}a) and Fig.(\ref{fig4}b) with temperature, 
$T=0.5\Delta_{0}$.
The color plots yield the information of the Seebeck coefficient for different values of the RSOC and 
the ISOC parameters.  For certain values of the RSOC parameter ($>0.1t_1$), irrespective
 of the ISOC strength, both spins show higher values of thermopower.
Further we have shown results for the charge and spin Seebeck coefficients 
in Fig.(\ref{fig4}c) and Fig.(\ref{fig4}d). 
The charge Seebeck coefficient shows higher values for larger strengths of RSOC, and for certain values of the SOC parameters, the spin Seebeck coefficient vanishes.
This map aids in identifying an operating regime of the magnitude of the Seebeck coefficient corresponding to a variety of choices of $\lambda_{R}$ and $\lambda_{I}$. As 
the strengths of SOCs  correspond to presence of different adatoms, a careful choice of the periodic table may
provide useful information on tunable thermopower of these junction devices.

\begin{figure}[!h]
\centering
\includegraphics[scale=.17,{angle=270}]{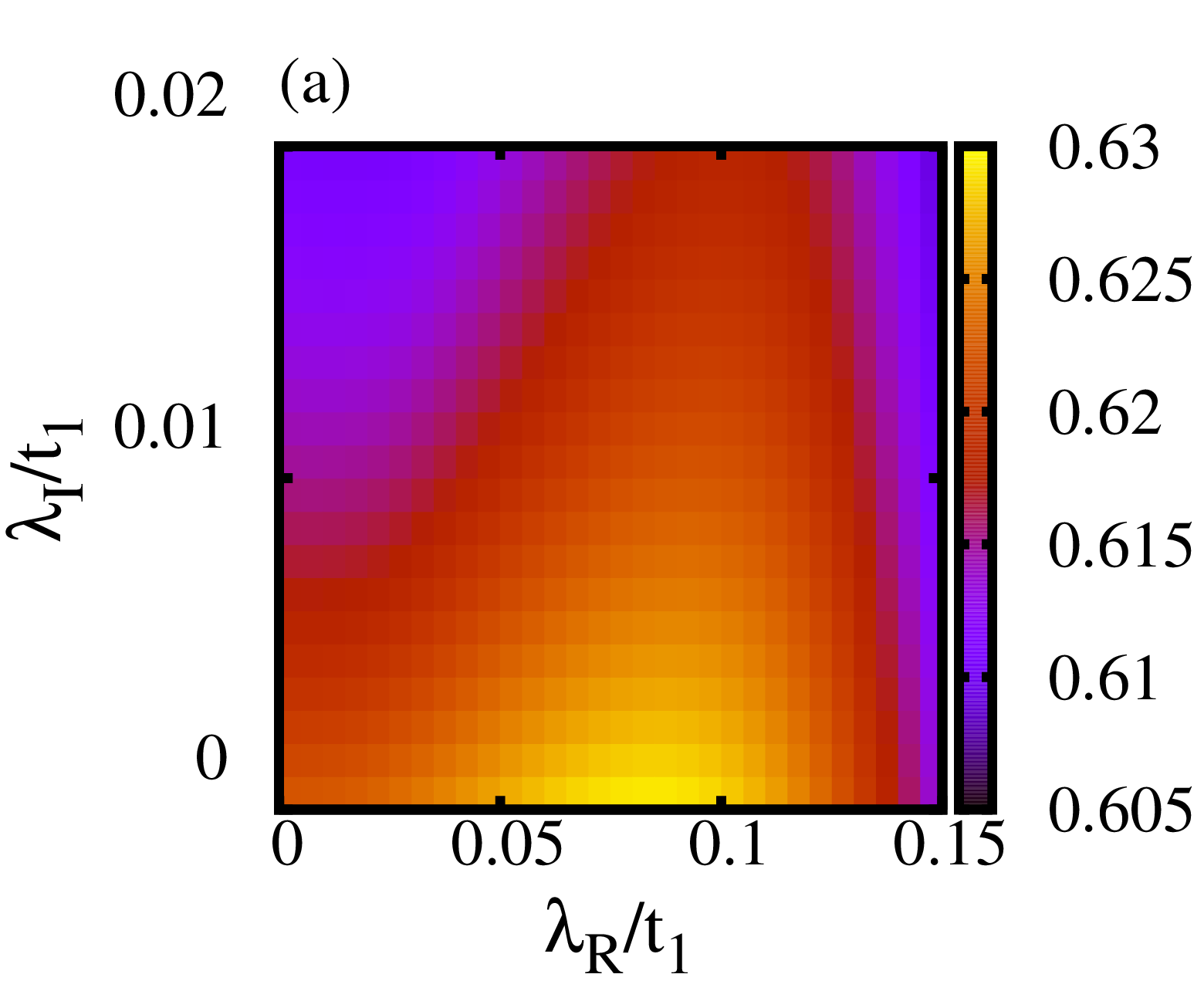}\hspace{0.01in}
\includegraphics[scale=.17,{angle=270}]{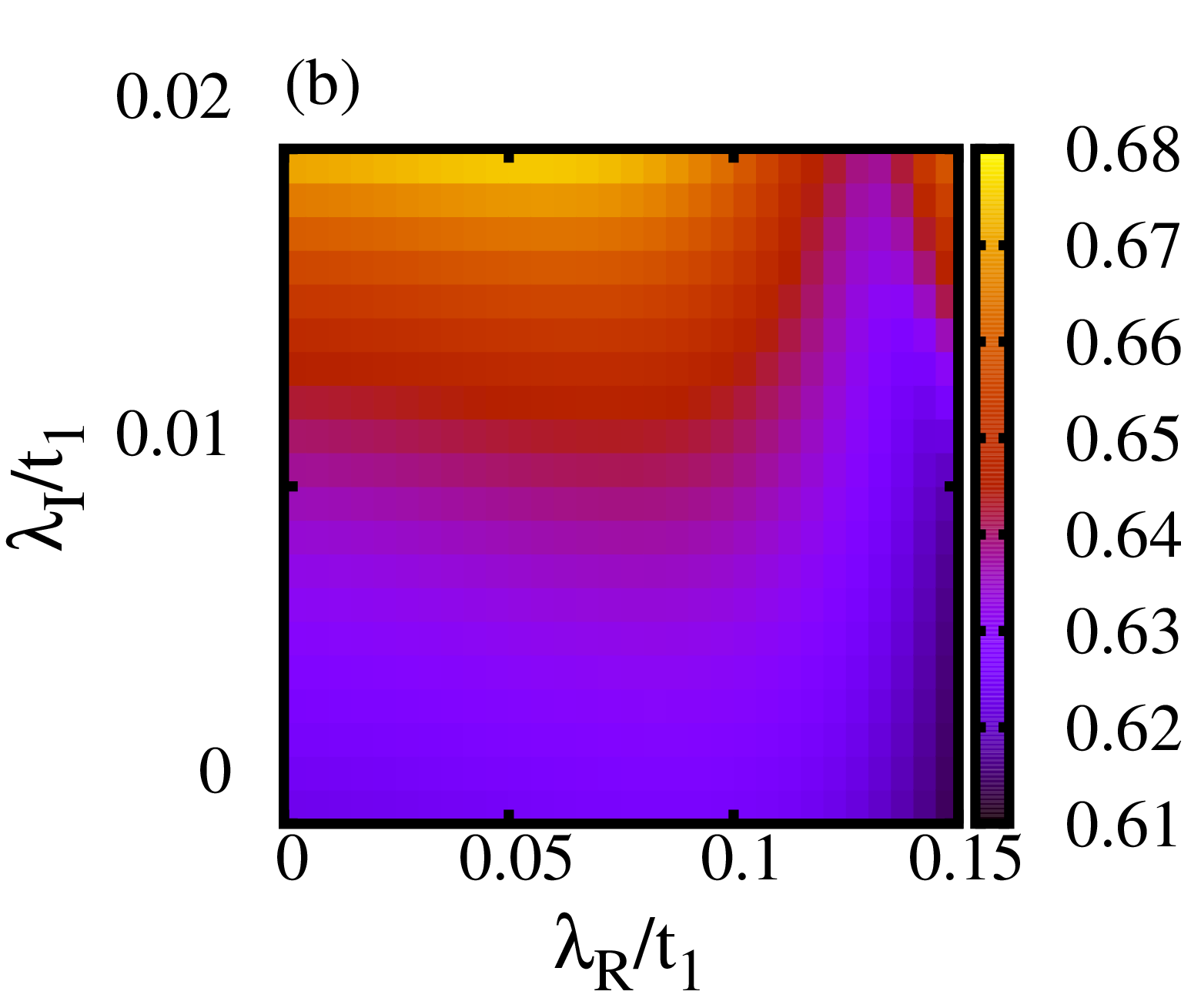}\\
\includegraphics[scale=.17,{angle=270}]{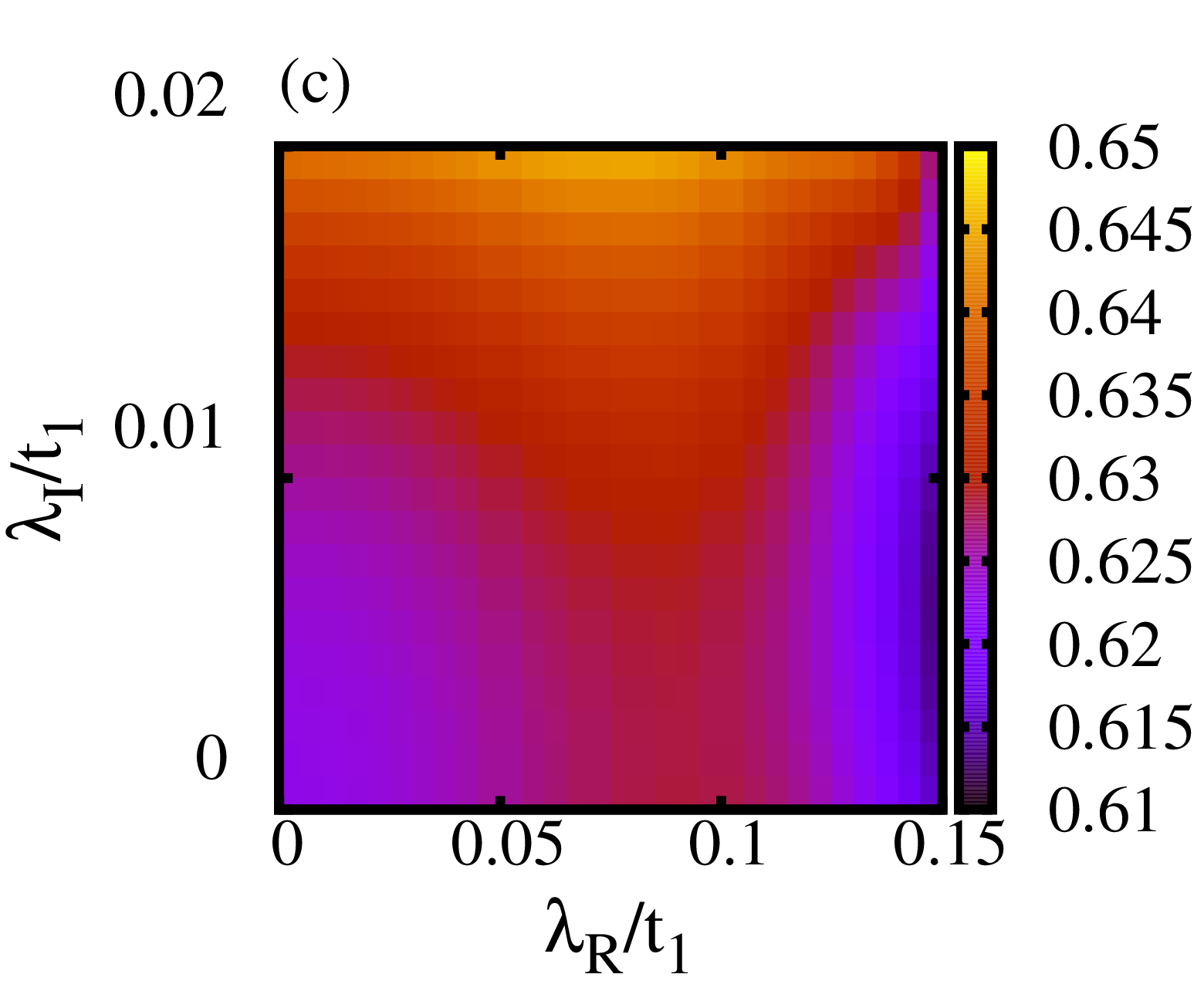}\hspace{0.01in}
\includegraphics[scale=.17,{angle=270}]{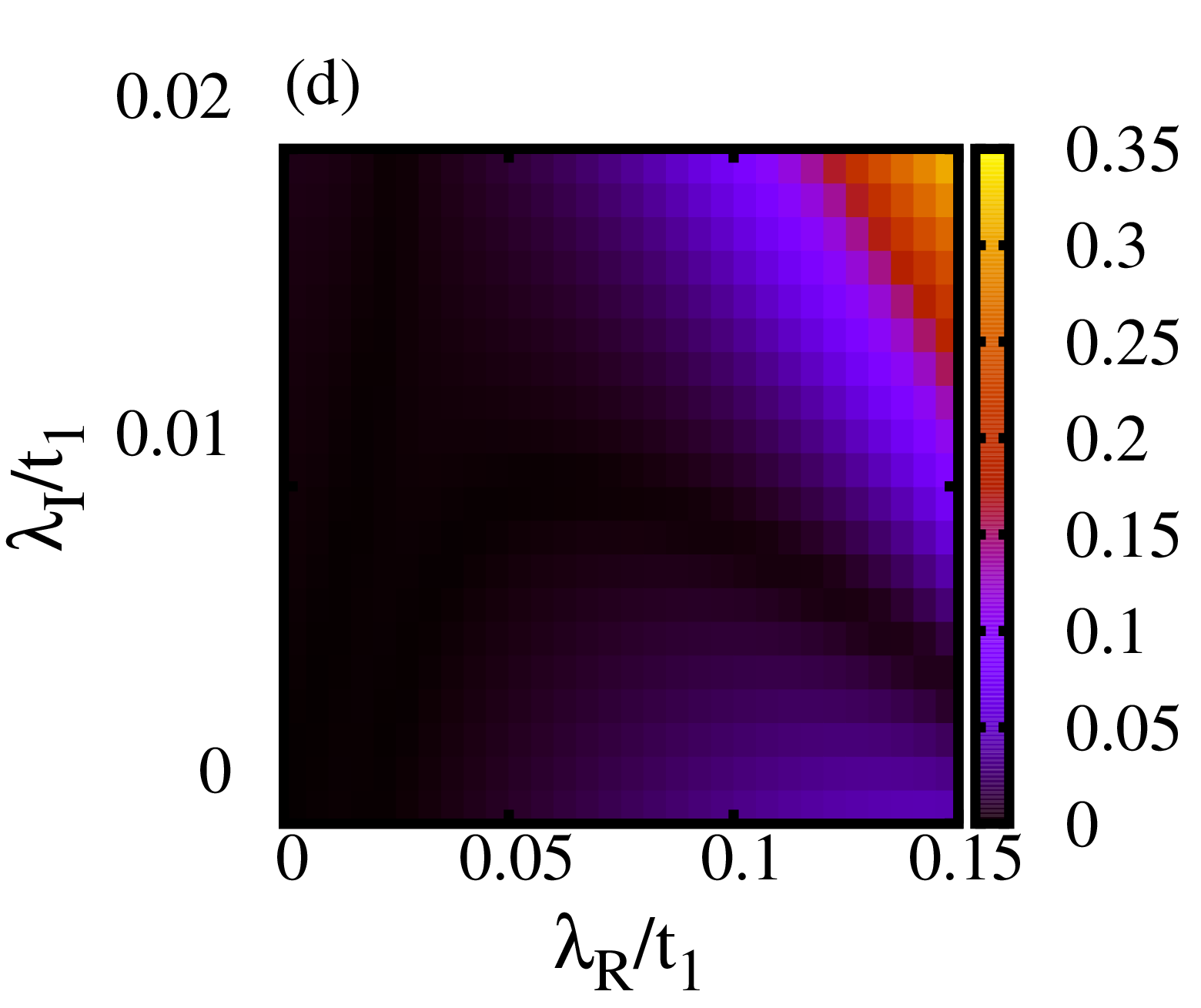}
\caption{(Color online)
	The variation of the spin resolved Figure of Merit, $Z_{\sigma}T$ as function of $\lambda_R$ and $\lambda_I$ (scaled by $t_1$) for
$(a)$ up spin,
$(b)$ down spin,
	$(c)$ The variation of the charge FM, $Z_{ch}T$ as function of $\lambda_R$ and $\lambda_I$ (scaled by $t_1$),
	$(d)$ The variation of the spin FM, $Z_{sp}T$ as function of $\lambda_R$ and $\lambda_I$ (scaled by $t_1$).}
\label{fig55}
\end{figure}

Now we show the results on the 'Figure of Merit' (FM) which defines the efficiency of this system as a thermopower device.
The variations of the spin dependent FM, $Z_{\sigma}T$ as the function of the 
spin-orbit couplings ($\lambda_R$ and $\lambda_I$) are presented in Fig.(\ref{fig55}a) and Fig.(\ref{fig55}b).
%It is clear that for both the spins, the FM and the Seebeck coefficient show their higher values for similar values of the SOC. 
Interestingly,
the down spin shows more efficiency compared to that of the up spin and hence it contradicts the behaviour of the Seebeck coefficient.   
Further we have shown the results for charge and spin FM in  Fig.(\ref{fig55}c) and Fig.(\ref{fig55}d),  where it is observed that for higher values of  ISOC
the charge FM becomes larger.  The spin FM becomes zero for the lower values of the RSOC parameters irrespective of the ISOC strength. Such regions, alongwith
others, are shown by black patches in Fig.(\ref{fig55}d).
Thus these maps aid in deciding on the values of the parameters that may be used for maximizing the gain of these KMNIS junction devices.  
%The variation of the thermal current as the function of RSOC parameter for two different values of effective barrier potential has
%been presented in Fig.(\ref{fig4}) where the driving voltage $V_{B}$ is fixed at $0.6\Delta_0$. 
%\begin{figure}[!h]
%\centering
%\includegraphics[scale=0.3]{2a.eps}\\
%\caption{(Color online) The red line with circles
%is for $\chi=10\eta/2$, the blue line with squares is for $\chi=11\eta/2$.
%The variation of $2J_{NS}e^{2}R_{N}/\Delta_{0}^{2}$  as function of RSOC strength.}
%  \label{fig4}
%\end{figure}  
%In particular, we have considered the values of $\chi$ as $10\eta/2$ and $11 \eta/2$ to show 
%the contrasting nature of the behaviour of thermal current in presence of RSOC. It is clear that RSOC enhances the thermal current 
%for the effective barrier potential to be an odd multiple of $\eta/2$, and for effective barrier potential to be an 
%even multiple, up to a certain strength of the RSOC parameter the reverse happens. For large values of the RSOC strength both the plots
%show a decreasing trend. Although tunable, RSOC rarely becomes very large, and thus the lower to moderate values of RSOC parameter is of experimental
%importance. We observe interesting Physics to be occurring precisely in this region.

%====================================================================================================================================
\subsection{ \label{subsec:3.2} Thermoelectric cooling and coefficient of performance}
\begin{figure}[!h]
\centering
\includegraphics[scale=0.17]{Fig6a.eps}\\\vspace{0.01in}
\includegraphics[scale=0.17,{angle=270}]{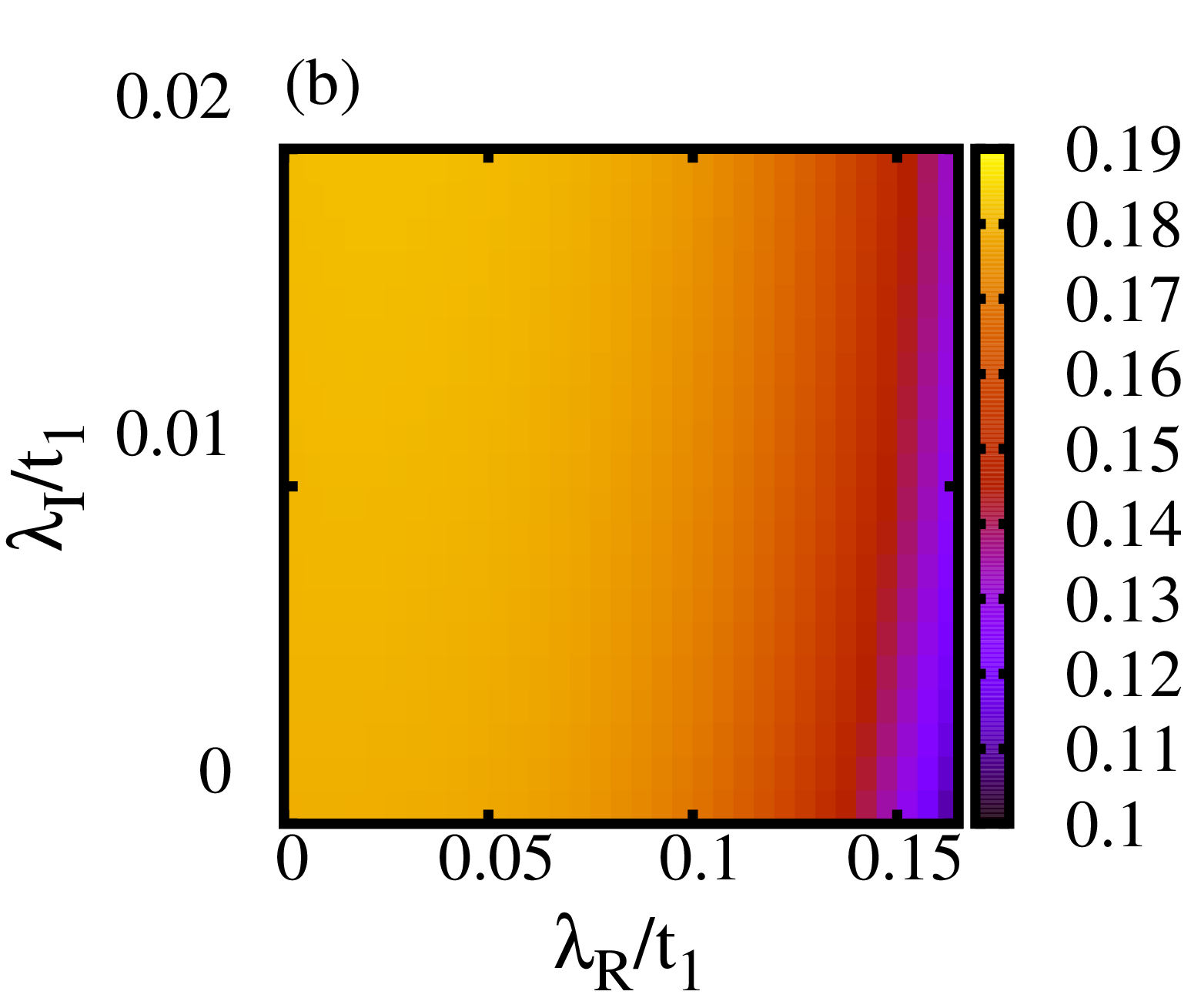}\hspace{0.01in}
\includegraphics[scale=0.17,{angle=270}]{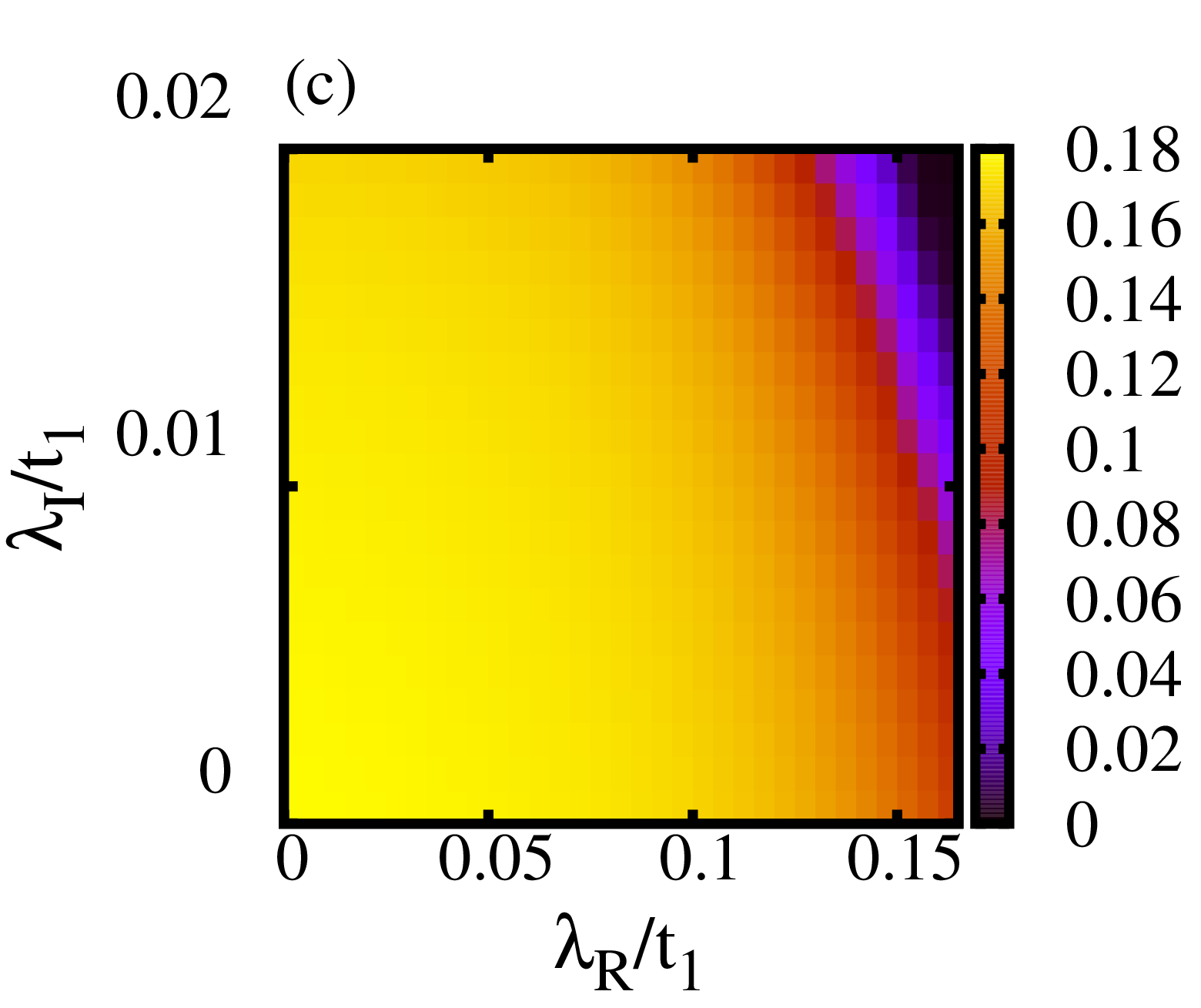}
\caption{(Color online)
$(a)$The variation of the thermal current, $2J_{NS}e^{2}R_{N}/\Delta_{0}^{2}$  as a function of biasing voltage, 
	$V_B$ (scaled by superconducting order parameter, $\Delta_{0}$).
The variation of the thermal current as function of $\lambda_R$ and $\lambda_I$ (scaled by $t_1$) for
$(b)$ up spin,
$(c)$ down spin.
}
  \label{fig6}
\end{figure}

Here we show the results of the thermoelectric cooling of the KMNIS junction. 
To get an idea about the role of spin-orbit coupling on the thermal current, in Fig.(\ref{fig6}a) we present the variation of a dimensionless quantity,
$2J_{NS}e^{2}R_{N}/\Delta_{0}^{2}$ as a function of biasing voltage where the temperature is fixed at $T=0.5\Delta_0$.
It is evident that at zero biasing voltage,
the rate of the thermal current extracted from the cold (normal) reservoir is negative. Thus, to
achieve cooling effects, a lower threshold voltage of the battery, namely, $V_{lower}$ is
needed. That is, the thermoelectric cooling does not work when $V_B < V_{lower}$. Also beyond 
an upper threshold voltage, $V_{upper}$, the refrigeration
effect ceases to exist. From the plots it is visible that for a very small range of the biasing voltage (having values
in the vicinity of the superconducting order parameter), the thermoelectric cooling process is
efficient and the maximum refrigeration occurs at around, $V_B\sim 0.9\Delta_0$.
Moreover,  we have checked that a lower threshold of the voltage, $V_{lower}$ required 
to trigger the cooling effect does not depend on the choice of the SOC parameters. 
%Further,
%we have checked that the graphene based NIS junction devices show comparatively low thermoelectric cooling with respect to that of the normal NIS junction devices.

Next we have shown  spin resolved thermal current as the function
of both the spin-orbit couplings in Fig.(\ref{fig6}b) and Fig.(\ref{fig6}c)  which provides
 an idea of the desired (for maximum thermoelectric cooling) range of values of the RSOC,
the ISOC parameters and the corresponding thermal current. The biasing voltage is fixed at $V_B=0.5\Delta_0$. 
 For higher values of the RSOC parameter, the thermoelectric cooling for both spins becomes smaller.
As the SOC parameters are tunable, the color plots gives useful information on
the tunable thermoelectric cooling.
%Let us take an example for illustration of the cooling effects. 
%Considering the cooling for biasing voltage $0.5\Delta_0$, one can achieve cooling of a metal by approximately $90K$ by taking the contribution for both the spins.
%(see footnote \footnote{Corresponding to $V_{B}=0.5\Delta_{0}$, $J_{NS}$ is $\sim 0.6$ in units of $\Delta_{0}^{2}/e^2R_{N}$ (considering both spin contribution). Multiplying $0.6$ by $\Delta_{0}^{2}$ 
%(with $\Delta_{0}=0.01eV$) and using $1K=8.62\times 10^{-5}eV$ one gets $90K$.}). 

Here we show the results of the performance of the KMNIS junction as a self-cooling device. To recapitulate, a measure of the performance and the efficiency of the KMNIS junction 
as a thermoelectric nano-refrigerator is defined by the
coefficient of performance ($COP$).
\begin{figure}[!h]
\centering
\includegraphics[scale=0.17]{Fig7a.eps}\\\vspace{0.01in}
\includegraphics[scale=0.17,{angle=270}]{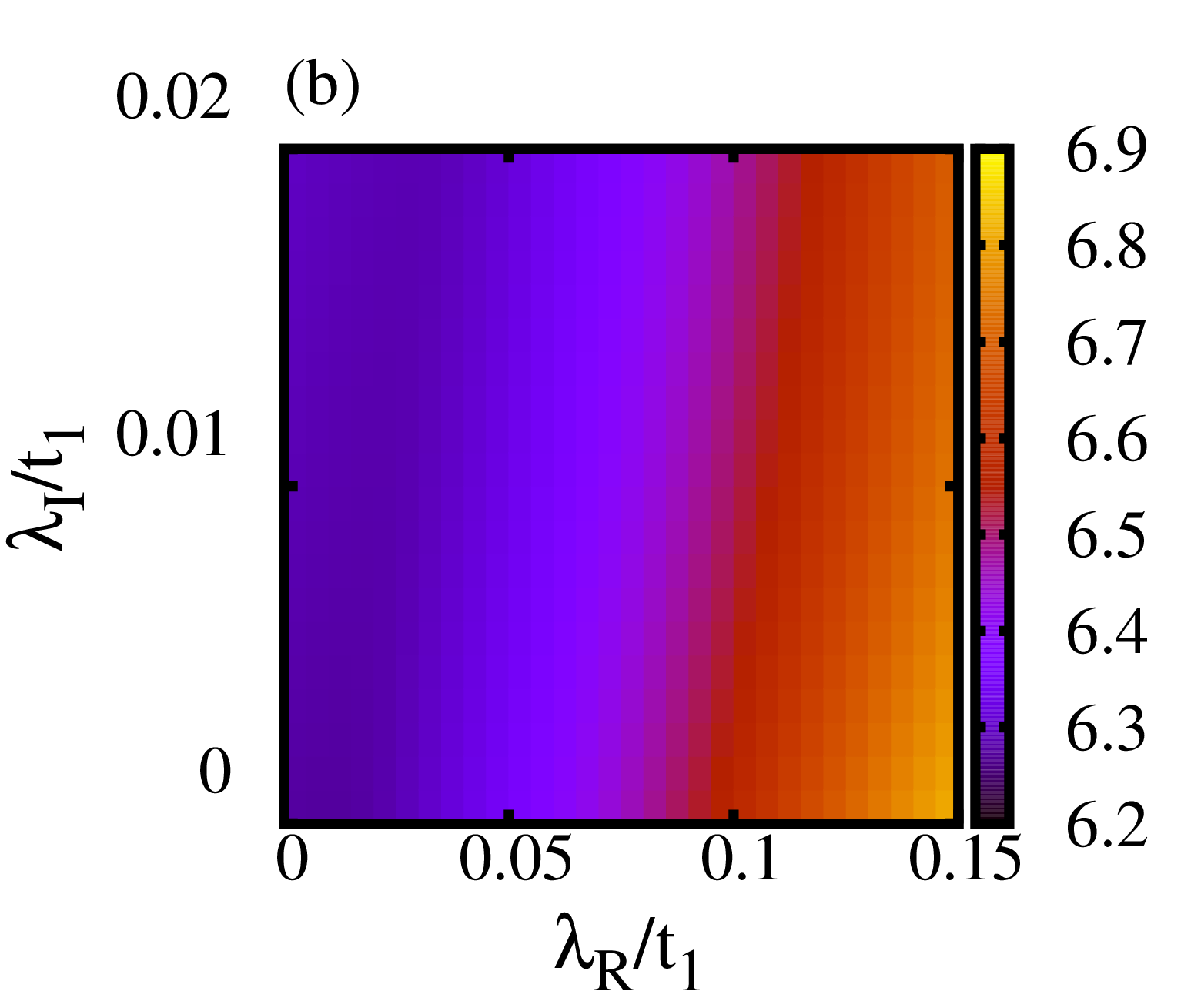}\hspace{0.01in}
\includegraphics[scale=0.17,{angle=270}]{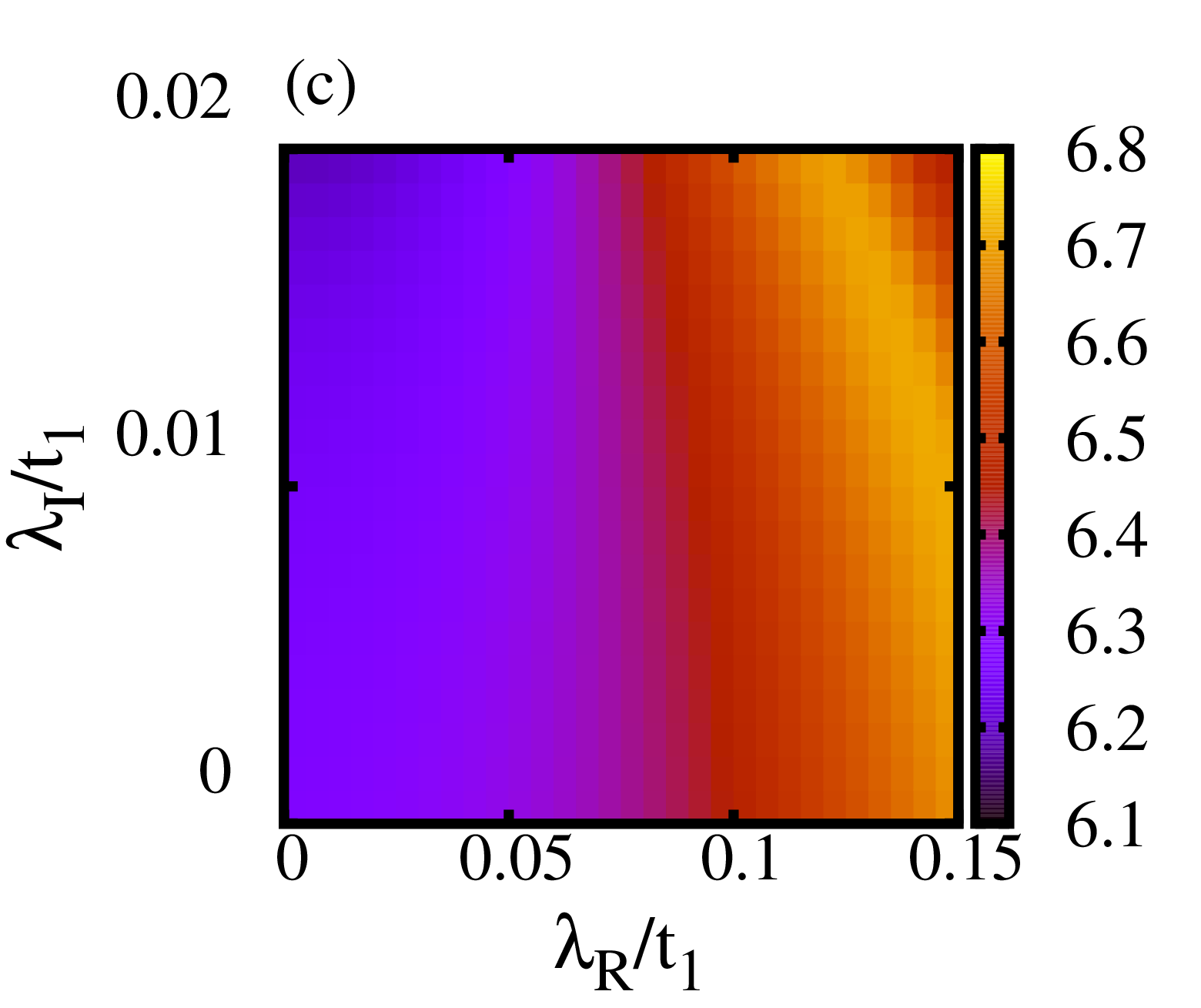}
\caption{(Color online)
$(a)$The variation of the coefficient of performance, $COP$  as a function of biasing voltage, $V_B$ (scaled by superconducting order parameter, $\Delta_{0}$).
The variation of $COP$ as function of $\lambda_R$ and $\lambda_I$ (scaled by $t_1$) for
$(b)$ up spin,
$(c)$ down spin.
}
  \label{fig7}
\end{figure}
Fig.(\ref{fig7}) reveals that for a {\it narrow} range of the biasing voltage, the nano refrigeration works (namely, $\sim 0.06\Delta_{0}-\Delta_{0}$). 
Since $\Delta_0$ is in $meV$ range for conventional superconductor, the operating voltage is low.
Further it is understood that when the driving voltage exceeds the value of the superconducting
order parameter, $\Delta_{0}$, the refrigeration vanishes, irrespective of the strengths of SOC.
%Further it is checked that the graphene based NIS junction shows more $COP$ compared to that of the normal NIS junction devices.

Finally to complete our enumeration of the tunability of a KMNIS junction,  the
$COP$ is plotted as a function of the RSOC and ISOC strengths in  Fig.(\ref{fig7}b) and Fig.(\ref{fig7}c).
The color plots
provide the idea of the range of values of the RSOC, the ISOC 
and the corresponding coefficient of performance. The biasing voltage is fixed at $V_B =0.125\Delta_0$ where the $COP$ shows large value.
It is clear that the RSOC helps to enhance the $COP$ for both spins.
The color plots provide a helpful hint on the
tunable coefficient of performance.

%Finally to complete our enumeration of the tunability of a MIS junction with regard to the refrigeration, the
% $COP$ is plotted as a function of the Rashba
%coupling strength, $U$ for two different values of $\chi$ in Fig.\ref{fig7} where the driving voltage, $V_{B}$ is fixed at $V_{B}=0.08\Delta_0$
%(where $COP$ is maximum as shown in Fig.(\ref{fig6})).
%\begin{figure}[!h]
%\centering
%\includegraphics[scale=0.3]{3b.eps}\\
%\caption{(Color online) The red line with circles
%is for $\chi_{1}$, the blue line with squares is for $\chi_{2}$.
%The variation of $COP$  as function of $U$.}
%  \label{fig7}
%\end{figure}  
%The $COP$ as the function of the RSOC parameter shows a different nature for two different 
%values of effective barrier potential, $\chi_{1}$ and $\chi_{2}$ corresponding to which the contrast in the behaviour of $COP$ can be 
%best illustrated. Thus
%the effective barrier potential reserves a right to decide on behaviour of $COP$ on the RSOC parameter. 
%The preceding discussion makes it amply clear that $U$ and $\chi$ which are assumably under control in an
%experimental setup, have critical roles to play in the phenomenon of nano refrigeration.  

%=================================================================================================================  
\section{Conclusion}
\label{sec:4}
In summary, we have investigated the
thermoelectric properties of a KMNIS junction in presence 
 of spin-orbit couplings by employing BTK theory in great details.
We have investigated the Seebeck coefficient and the thermoelectric Figure of Merit of this junction device
and explained how SOC play a role in determining these properties. Further the thermoelectric cooling of this junction device
and its performance as a cooling device have been studied in details. It is clear that the thermoelectric properties are sensitive to
the SOC parameters.
As the strength of the SOC terms can be
manipulated via different adatoms or using gate voltages
etc, we may infer that it is possible to achieve a precision tuning of the thermoelectric properties of these junction devices.
%\acknowledgments
% SB thanks SERB, India for financial support under the grant
%F. No: EMR/2015/001039.

\end{document}